\documentclass[letterpaper]{article} %
\usepackage{aaai25}  %
\usepackage{times}  %
\usepackage{helvet}  %
\usepackage{courier}  %
\usepackage[hyphens]{url}  %
\usepackage{graphicx} %
\usepackage{natbib}  %
\usepackage{caption} %
\usepackage{algorithm}
\usepackage{algorithmic}

\usepackage{newfloat}
\usepackage{listings}
\DeclareCaptionStyle{ruled}{labelfont=normalfont,labelsep=colon,strut=off} %
\floatstyle{ruled}
\newfloat{listing}{tb}{lst}{}
\floatname{listing}{Listing}
\usepackage{booktabs}
\usepackage{multirow}
\usepackage{makecell}
\usepackage{amsmath}
\title{Personalized Dynamic Music Emotion Recognition \\ with Dual-Scale Attention-Based Meta-Learning}

\author {
    Dengming Zhang\textsuperscript{\rm 1},
    Weitao You\textsuperscript{\rm 2}\thanks{Corresponding author},
    Ziheng Liu\textsuperscript{\rm 2},
    Lingyun Sun\textsuperscript{\rm 2},
    Pei Chen\textsuperscript{\rm 2}
}

\affiliations {
    \textsuperscript{\rm 1}School of Software Technology, Zhejiang University, China\\
    \textsuperscript{\rm 2}College of Computer Science and Technology, Zhejiang University, China\\
    \{dmz, weitao\_you, northstar\_lzh, sunly, chenpei\}@zju.edu.cn
}

\usepackage{bibentry}

\begin{document}

\maketitle

\begin{abstract}
Dynamic Music Emotion Recognition (DMER) aims to predict the emotion of different moments in music, playing a crucial role in music information retrieval. The existing DMER methods struggle to capture long-term dependencies when dealing with sequence data, which limits their performance. Furthermore, these methods often overlook the influence of individual differences on emotion perception, even though everyone has their own personalized emotional perception in the real world. Motivated by these issues, we explore more effective sequence processing methods and introduce the Personalized DMER (PDMER) problem, which requires models to predict emotions that align with personalized perception. Specifically, we propose a Dual-Scale Attention-Based Meta-Learning (DSAML) method. This method fuses features from a dual-scale feature extractor and captures both short and long-term dependencies using a dual-scale attention transformer, improving the performance in traditional DMER. To achieve PDMER, we design a novel task construction strategy that divides tasks by annotators. Samples in a task are annotated by the same annotator, ensuring consistent perception. Leveraging this strategy alongside meta-learning, DSAML can predict personalized perception of emotions with just one personalized annotation sample. Our objective and subjective experiments demonstrate that our method can achieve state-of-the-art performance in both traditional DMER and PDMER.
\end{abstract}

\begin{links}
    \link{Code \& Case}{https://littleor.github.io/PDMER}
\end{links}

\section{Introduction}
Music Emotion Recognition (MER) technology focuses on identifying emotions conveyed by music, applying to music therapy \cite{dingle2015influence}, music recommendation \cite{liu2023emotion, tran2023emotion}, and music generation \cite{huang2020emotion, ji2024muser}.
To describe the music emotion, Russell's two-dimensional valence-arousal (V-A) emotional model \cite{russell1980circumplex} is widely used in MER, where valence describes the extent to which an emotion is positive or negative, and arousal refers to its intensity.
 Existing MER tasks are generally divided into static MER (SMER) and dynamic MER (DMER) \cite{han2022survey}.
 SMER inputs music and outputs only one V-A label to describe the emotion, which fails to describe the variations in emotion within the music. For example, Beethoven's Symphony No. 5 can't be described as simply positive and intense, as it also contains moments of sadness and tranquility. In contrast, DMER predicts the V-A label sequence, using a sequence label to describe the emotional changes in the music, which can more accurately express the emotions.

Existing DMER work focuses on utilizing sequential information to predict, as the emotion of each moment in music is related to the emotions before and after. Specifically, long short-term memory (LSTM) \cite{hochreiter1997long} attracted the attention of researchers due to its superiority in sequence data processing. These researchers use LSTM to extract sequence features and have shown some effectiveness in DMER \cite{he2020multi, zhang2023dual}.
However, previous studies \cite{khandelwal2018sharp, li2019enhancing, grigsby2021long} have found that LSTM struggles to capture long-term dependencies, which limits the capture of the global emotion of music, resulting in poor performance in DMER.

\begin{figure}[t]
\centering
\includegraphics[width=\columnwidth]{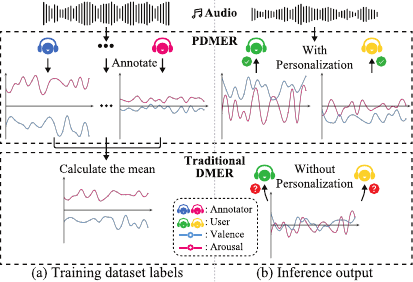} %
\caption{
    The differences between traditional DMER and PDMER.
    All charts represent the emotion valence/arousal (V/A) curve of music, where the x-axis represents time and the y-axis represents V/A values.
}
\label{fig:PDMER}
\end{figure}

More importantly, existing work often assumes that all individuals perceive music emotions in the same way, neglecting the significant impact of individual differences on DMER. For example, MER1101 dataset \cite{zhang2023dual} involves multiple annotators for labeling and uses their average as labels to attempt to eliminate individual differences in the data as shown in Figure \ref{fig:PDMER}(a).
However, emotions are personalized, and different individuals have different perceptions of emotions towards the same song \cite{kang2024we}. For example, a song that makes one person feel happy may make another person feel sad.
These existing works eliminates individual differences in data while eliminating bias, using group emotional perceptions instead of individual emotional perceptions to avoid personalization issues.
To make matters worse, real-world applications often face challenges with diverse emotional perceptions from each individual, and removing personalization differences in datasets cannot address this issue. Therefore, we point out the personalized DMER (PDMER) problem, which requires models to predict emotions that align with individual personalized perceptions rather than group perceptions.

To address the above issues, we propose a Dual-Scale Attention-Based Meta-Learning (DSAML) method to handle PDMER tasks. Specifically, to achieve DMER, DSAML uses a pre-trained Imagebind model \cite{girdhar2023imagebind} to extract global audio features and introduces an adapter to extract trainable local audio features. Fusion of the two features is fed into the dual-scale attention transformer, which focus on both local and global features to capture more comprehensive emotional features. Finally, a sequence of V-A values is predicted by the sequence prediction module, completing the DMER.
To achieve PDMER, the Model-Agnostic Meta-Learning (MAML) \cite{finn2017model} is used to personalize the model. Furthermore, we propose a personalized meta-learning task construction strategy for MAML, which divided tasks by annotators instead of music samples. Samples in a task are annotated by the same annotator, ensuring consistent perception. With this strategy, DSAML preserves personality differences, effectively enhancing PDMER performance.

In summary, this paper has the following contributions:
\begin{itemize}
    \item
    We propose the DSAML including a dual-scale feature extractor and a dual-scale attention transformer to capture both local and global features, improving the performance of traditional DMER.
    \item We recognize the importance of personalized emotional perception for DMER, introduces the PDMER task, and design a personalized prediction method based on meta-learning with a novel task construction strategy.
    \item Objective experiments demonstrate that our method achieves the best performance in both traditional DEMR and PDMER. Subjective experiments also show that our method better conforms to individual personalized emotional perception in the real world.
\end{itemize}

\section{Related Work}
    \subsection{Dynamic Music Emotion Recognition}
    DMER aims to predict the emotions of music at different moments, as the emotions at any moment are often related to those before and after, so it is necessary to consider the temporal dependencies in the music emotions.
    As the recurrent neural network (RNN) is suitable for sequence data processing, researchers first used RNN to extract sequence features for DMER \cite{malik2017stacked}.
    However, due to the issue of gradient vanishing in RNNs when dealing with long sequences, researchers in the DMER field have subsequently adopted LSTM to extract sequential features, as LSTM introduces gating mechanisms for long sequence processing.
    Specifically, Zhang et al. \cite{zhang2023dual} integrated spatial and channel dimension features and used Bi-directional LSTM (BiLSTM) for sequence learning to predict the V-A sequence of music. He et al. \cite{he2020multi} used multi-view CNN as feature extractors and then used BiLSTM to capture temporal context for predicting the V-A sequence of music. These methods have shown some effectiveness in DMER, but they still struggle to capture long-term dependencies as previously studies demenstrated \cite{khandelwal2018sharp, li2019enhancing,
    grigsby2021long}.
    Therefore, we propose a dual-scale feature extractor and a dual-scale attention transformer to capture both short- and long-term dependencies, thereby improving the performance of long-sequence data processing.

    \subsection{Personalized Music Emotion Recognition}
    In the SMER field, researchers have conducted some PMER studies, Yang et al. \cite{yang2007music} first quantitatively evaluated the impact of personality on MER and found that personalization significantly influences MER. Based on this quantitative experiment, numerous personalized MER studies have appeared in the field of SMER.
    To achieve PMER, researchers mainly train personalized models using samples annotated by specific users. Personalized training depends on a large amount of user-annotated data, so researchers focus on reducing the number of specific user-annotated samples. Su and Fung \cite{su2012personalized} proposed an active learning method to reduce the number of specific user-annotated samples by selecting the most informative ones for manual annotation. Wang et al. \cite{wang2012personalized} and Chen et al. \cite{chen2014linear} used two-stage training, first training the background model and then adapting it using fewer specific user-annotated samples.

    However, in the field of DMER, there has been no specialized research on PMER, researchers often use the mean of multiple data annotators as the emotion label which ignores the impact of individual personalization on MER \cite{orjesek2019dnn, zhang2023dual}. In other words, many pieces of DMER research focus on group emotions rather than individual emotions, which limits the practical application performance of existing DMER.
    Although the SMER methods have achieved good results in PMER, they still require a large amount of user-annotated data (at least 20 samples) \cite{chen2014linear}. In DMER, users need to annotate more labels (e.g., 1200 labels for 20 30-second music, assuming labeling every 0.5 seconds), making these personalized SMER methods difficult to apply in DMER. Therefore, we design a PDMER method based on meta-learning with a new meta-learning task construction strategy, which only requires one specific user-annotated sample.

    \subsection{Meta-learning}
    Meta-learning is widely used in few-shot learning problems because it can learn to solve new tasks with a small number of samples \cite{thrun1998learning, wang2020generalizing}. Meta-learning methods can be generally divided into three types: metric-based, model-based, and optimization-based meta-learning methods. Among them, metric-based meta-learning methods focus on solving classification problems in the feature space \cite{snell2017prototypical, sung2018learning} and are difficult to apply to regression tasks like DMER; model-based meta-learning methods focus on designing specific model structures to achieve fast learning goals \cite{wang2020generalizing}, which limits the model structure; while optimization-based meta-learning methods adjust existing optimization algorithms to converge with a few samples for new tasks \cite{finn2017model, nichol2018first}. For example, MAML has no constraints on model structure and can also be applied to regression tasks, making it most suitable for application in DMER. However, directly applying MAML in PDMER still ignores individual personalized effects during the training process, limiting the performance of MAML. Therefore, we proposes a new method for constructing meta-learning tasks: incorporating the annotated data of each annotator and constructing tasks according to annotators, thereby effectively improving the performance of the model in the PDMER task.

\section{Methods}

    \subsection{Problem Formulation} \label{sub_sec:problem_formulation}
    Given a training dataset $ \mathbf{D} = \{(x_1, y^1_1), \ldots, (x_i, y^j_i), \ldots\} $, and personalized annotated data (annotator not present in the training set) $ \mathbf{S}_{p} = \{(x_{p1}, y^p_{p1}), \ldots, (x_{pn}, y^p_{pn})\} $, where $ x_i $ is the $ i $-th music sample in $\mathbf{D}$, $ y^j_i $ is the label from the $ j $-th annotator for $ x_i $, $p$ represents any user with personalized emotional perception, $x_{pn}$ is the $ n $-th music sample in $ \mathbf{S}_p $, and $y^p_{pn}$ is the label for $x_{pn}$ from the user $p$.
    In traditional DMER, the goal is to directly predict non-personalized label $y_q$ for a new query sample $ q $ by training on $\mathbf{D}$.
    However, in PDMER, our goal is to predict the personalized label $y^p_q$ for the specific user $p$ by training on $\mathbf{D}$ and adapting it with $\mathbf{S}_p$.

    \subsection{Model Architecture}

    \begin{figure*}[!h]
    \centering
    \includegraphics[width=\textwidth]{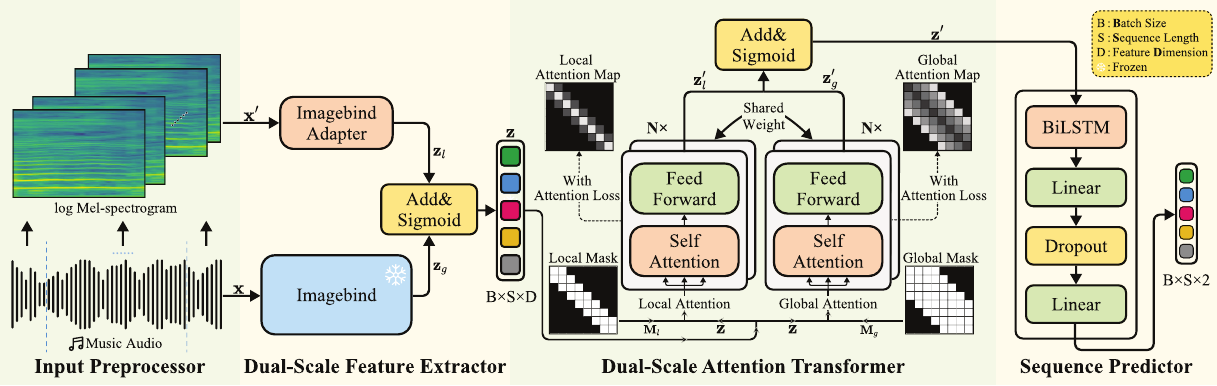} %
    \caption{The architecture of the DSAML model.}
    \label{fig:model_architecture}
    \end{figure*}

    The DSAML model consists of four parts: Input Preprocessor, Dual-Scale Feature Extractor, Dual-Scale Attention Transformer, and Sequence Predictor, as shown in Figure \ref{fig:model_architecture}.

    \subsubsection{Input Preprocessor}
    To achieve DMER, the model needs to extract sequence features from the audio. Therefore, in the input preprocessor, we slice the original audio input. A music segment of length $ l $ seconds is sliced into $ k $ segments, corresponding to $ k $ sequence prediction values and a resolution of $ k / l $ Hz. The sliced music segments are then processed to calculate the log Mel-spectrogram.

    \subsubsection{Dual-Scale Feature Extractor}
\begin{figure}[!h]
    \centering
    \includegraphics[width=\columnwidth]{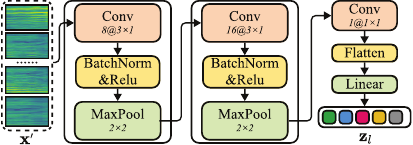} %
    \caption{The architecture of the Imagebind Adapter}
    \label{fig:adapter}
    \end{figure}

    The pre-trained Imagebind model has shown good performance in audio feature extraction \cite{zou2023emid, chakhtouna2024modeling}. Consequently, DSAML utilizes Imagebind to extract the global feature $ \mathbf{z_g} $ of the audio $ \mathbf{x} $. However, the large number of parameters in Imagebind would significantly increases the personalization adaptation time in PDMER. Therefore, this paper freezes the parameters of Imagebind and introduces the Imagebind Adapter module with significantly fewer parameters. Furthermore, since the global feature extracted by Imagebind cannot represent the finer emotional changes in music, the Imagebind Adapter is designed to extract the local feature $ \mathbf{z_l} $ from the spectrogram sequence $ \mathbf{x}^{\prime} $ of the short music segments. The local and global features are then fused to obtain the audio feature $\mathbf{z} = \sigma (\mathbf{z}_l + \mathbf{z}_g)$, where $ \sigma $ refers to the Sigmoid function. Specifically, as shown in Figure \ref{fig:adapter}, the Imagebind Adapter module extracts features using two convolutional layers, reduces the number of channels to 1 through a $ 1 \times 1 $ convolutional layer, and finally maps the feature dimension to $ D $ through a fully connected layer.

    \subsubsection{Dual-Scale Attention Transformer}
    Considering that the emotional state of music at a particular moment is often related to the music in the preceding moments, and that the overall emotion also influences the emotional state at that moment, this paper proposes a Dual Attention Transformer to extract context-aware emotional features $\mathbf{z}^{\prime}$ from audio features $\mathbf{z}$. The Dual Attention Transformer achieves different scales of attention through a local mask $\mathbf{M}_l$ and a global mask $\mathbf{M}_g$, where the context length $n_l$ of $\mathbf{M}_l$ is much smaller than the context length $n_g$ of $\mathbf{M}_g$, thereby enabling the Transformer’s dual attention to focus on both local and global emotions. For a mask $\mathbf{M}_n$ with context length $n$, it can be expressed as:
    \begin{equation}
    \mathbf{M}_n[i, j] = \left\{ \begin{array}{ll} 1 & \text{if } |i - j| \leq n \\ 0 & \text{otherwise} \end{array} \right.
    \end{equation}
    where $i$ and $j$ represent different time steps in the sequence, and $|i - j| \leq n$ indicates that at any time step $i$, the information at time step $j$ can be seen only if their distance does not exceed $n$. Using $\mathbf{M}_l$ and $\mathbf{M}_g$ as masks, local features $\mathbf{z}_l^{\prime}$ and global emotional features $\mathbf{z}_g^{\prime}$ are extracted from the audio features $\mathbf{z}$:
    \begin{equation}
    \mathbf{z}_l^{\prime}=f_\theta(\mathbf{z}, \mathbf{M}_l);
    \mathbf{z}_g^{\prime}=f_\theta(\mathbf{z}, \mathbf{M}_g)
    \end{equation}
    where $f$ represents the transformer module, and $\theta$ represents the parameters of the transformer module. It is noteworthy that $\mathbf{z}_l^{\prime}$ and $\mathbf{z}_g^{\prime}$ are extracted using a transformer module with shared parameters, only utilizing different masks to extract features at different scales. Finally, the local and global emotional features are fused $\mathbf{z}^{\prime}=\sigma (\mathbf{z}_l^{\prime} + \mathbf{z}_g^{\prime})$ to obtain the context-aware emotional features $\mathbf{z}^{\prime}$.

\begin{figure}[!h]
        \centering
        \includegraphics[width=\columnwidth]{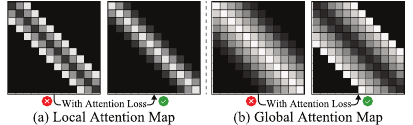}
        \caption{Attention Map}
        \label{fig:attention_map}
    \end{figure}

    However, our analysis of the local attention maps $\mathbf{A}_l$ and global attention maps $\mathbf{A}_g$ under different masks reveals that merely constraining the attention through masks does not ensure that $\mathbf{A}_l$ focuses more on the local context and $\mathbf{A}_g$ on the global context. As shown in Figure \ref{fig:attention_map}, it is possible that $\mathbf{A}_l$ might focus on distant moments and $\mathbf{A}_g$ on nearby moments, which results in extracted and fused features that are no longer comprehensive. To address this issue, we propose a diagonal attention map loss, which constrains the diagonal attention of $\mathbf{A}_l$ to be higher and that of $\mathbf{A}_g$ to be lower, ensuring that the two types of attention focus on different scales. It can be expressed as:
    \begin{equation}
    \small
    \begin{split}
        \mathcal{L}_{\text{attention}} = &\frac{1}{n} \sum_{i=1}^n [ (\text{diag}(\mathbf{A}_l)_i - \alpha)^2  + (\text{diag}(\mathbf{A}_g)_i - \beta)^2 ]
    \end{split}
    \end{equation}
    where $\text{diag}(\mathbf{A})$ denotes the diagonal values of the attention map $\mathbf{A}$, and $\alpha$ and $\beta$ are hyperparameters that control the diagonal attention of $\mathbf{A}_l$ and $\mathbf{A}_g$, respectively. Specifically, $\alpha$ is set to a higher value to ensure that $\mathbf{A}_l$ focuses more on the local context, while $\beta$ is set to a lower value to ensure that $\mathbf{A}_g$ focuses more on the global context.

    \subsubsection{Sequence Predictor}
    In the sequence predictor, DSAML further processes the sequence features using BiLSTM and reduces the feature dimension to 2 through fully connected layers to achieve regression of the V-A values, thus completing the DMER of the music.

    \subsection{Personalized Strategy}
    DSAML employs MAML for personalized learning and proposes a novel meta-learning task construction strategy to enhance the model's personalized prediction performance. The objective of MAML is to find a model parameter $\theta$ for the task distribution $p(\mathcal{T})$, such that the loss $\mathcal{L}_{\tau_i}$ is minimized after $k$ steps of learning on a randomly sampled task $\tau_i \sim p(\mathcal{T})$, which can be expressed as:
    \begin{equation}
    \underset{\theta}{\text{min}} \sum_{\tau_i \sim p(\mathcal{T})} \mathcal{L}_{\mathbf{Q}_i}[{U^k_{\mathbf{S}_i}(\theta)}]
    \end{equation}
    where $\mathbf{S}_i$ and $\mathbf{Q}_i$ are the support set and query set randomly sampled from $\tau_i$ ($\mathbf{S}_i \cap \mathbf{Q}_i = \emptyset$), and $U^k_{\mathbf{S}_i}(\theta)$ is the operator that updates $\theta$ $k$ times using the $\mathbf{S}_i$.

    Defined in the problem formulation, we have a training dataset  $ \mathbf{D} = \{(x_1, y^1_1), \ldots, (x_i, y^j_i), \ldots\} $ to find the optimal $\theta$, where each music has labels from multiple annotators.
    As shown in Figure \ref{fig:PDMER}(a), the existing DMER approaches often use the mean of all annotator labels for each music as the label, and train on the processed dataset $\mathbf{D}^\prime = \{(x_1, y^\prime_1), \ldots, (x_n, y^\prime_n)\}$, where $y^\prime_n = \frac{1}{N} \sum_{i=1}^{N} y^i_n$. The traditional task construction strategy for training with MAML using $\mathbf{D}^\prime$ can be represented as:
    \begin{equation}
    \small
    \mathbf{S}_i=\text{RandomSample}(\mathbf{D}^{\prime});\mathbf{Q}_i =\text{RandomSample}(\mathbf{D}^{\prime} - \mathbf{S}_i)
    \end{equation}

    Although this method performs well on traditional DMER and many datasets often directly provide $\mathbf{D}^\prime$ \cite{aljanaki2017developing, zhang2018pmemo}, this type of dataset and task construction approach loses personalized emotional preference information and essentially represents group emotional preferences. Therefore, we propose a novel meta-learning task construction strategy, which builds tasks directly from the dataset $\mathbf{D}$ based on the annotators:
    \begin{equation}
    \small
    \mathbf{S}_i=\text{RandomSample}(\mathbf{D}_{i});\mathbf{Q}_i =\text{RandomSample}(\mathbf{D}_{i} - \mathbf{S}_i)
    \end{equation}
    where $\mathbf{D}_i$ refers to all the music annotated by the $i$-th user, i.e., $\mathbf{D}_i=\{(x_1^i, y_1^i), (x_2^i, y_2^i), \dots, (x_m^i, y_m^i) \}$. This method treats each user's personality as different tasks, enabling the model to find a $\theta$ that can adapt quickly and perform optimally across all personalities.

    \subsection{Training \& Inference Process}
    During the training process, we construct task distribution $p(\mathcal{T})$ from the training dataset $\mathbf{D}$ using the above personalized task construction strategy.
    A batch of tasks is randomly sampled from $p(\mathcal{T})$, and from each task $\tau_i$, we will randomly sample support set $\mathbf{S}_i$ and query set $\mathbf{Q}_i$. Specifically, $\mathbf{S}_i$ is used to optimize $\theta$ and obtain $\theta^\prime_i = U^k_{\mathbf{S}_i}(\theta)$, then the training loss will be calculated on $\mathbf{Q}_i$.
    After accumulating the losses of all tasks in this batch, the model parameters will be updated as $ \theta \leftarrow \theta - \eta \nabla_{\theta}  \sum_{\mathcal{T}_i \sim p(\mathcal{T})} \mathcal{L}_{\mathbf{Q}_i}(\theta^\prime_i) $, where $\eta$ is the learning rate.
    This process repeats until the model converges, and the fitted parameters $\hat{\theta}$ are used for inference.

    In the inference process, we use all samples annotated by a new user $p$ (anyone absent from $\mathbf{D}$) in the personalized data $\mathbf{S}_p$. These samples serve as a support set to fast adapt the model $\hat{\theta}$ to the user's personalized emotional perception and obtain the personalized model $\hat{\theta}^\prime_p$ for the new user $p$. Finally, the model $\hat{\theta}^\prime_p$ can predict any music's emotion that aligns with the personalized perception of the new user $p$.

    \subsection{Implementation Details}
    The resolution of DSAML is 2Hz, indicating there is one label every 0.5 seconds. The model architecture consists of 3 layers of Transformer, with a mask context length of $n_l=5$ and $n_g=30$. In attention loss, $\alpha=0.5$ and $\beta=0.05$. During training, only one sample is used for fast adaptation (i.e., both $\mathbf{S}_i$ and $\mathbf{S}_p$ only contain 1 sample), and 15 samples are used for evaluation (i.e., $\mathbf{Q}_i$ contains 15 samples). The Adam optimizer \cite{kingma2014adam} is employed with a learning rate of 0.00005. We train the model for 2000 episodes on a single NVIDIA GeForce RTX 4090 GPU.

\section{Experiments}
    \subsection{Experiment Settings}

        \subsubsection{Dataset}
        The performance of DSAML is evaluated using two publicly available DMER datasets, both of which provide V-A value annotations every 0.5 seconds, with all unstable annotations from the beginning to 15 seconds removed.

\begin{figure}[t]
        \centering
        \includegraphics[width=\columnwidth]{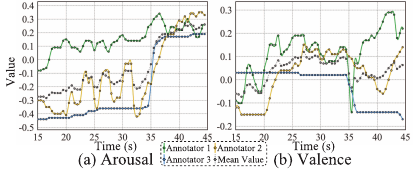}
        \caption{
            Example of personalized annotations for the same song with three different annotators in the DEAM dataset.
        }
        \label{fig:dataset_example}
        \end{figure}

        The first dataset is the DEAM dataset \cite{aljanaki2017developing}, which includes 1744 45-second clips and 58 full-length songs with an average length of 4 minutes, all containing dynamic annotations from each annotator  as illustrated in Figure \ref{fig:dataset_example}. In our experiments, we use the 58 full-length songs as the test set, with the remaining 1744 songs as the training set. Notably, 744 of the 45-second clips do not have annotator IDs, meaning we cannot determine the annotators for these songs. Therefore, in our proposed personalized task construction strategy, we only use 1000 songs as the training set.

        The second dataset is the PMEmo dataset \cite{zhang2018pmemo}, which includes 794 songs of varying lengths and the mean and standard deviation of the dynamic annotations, without individual annotator data. Thus, our personalized task construction strategy cannot be applied to this dataset, and this dataset is only used for evaluating traditional DMER tasks. We discard 122 samples with song lengths less than 25 seconds and use 40 songs longer than 65 seconds as the test set, with the remaining 632 songs as the training set.

        \subsubsection{Compared Models}
        Owing to the lack of research specifically targeting PDMER, this paper utilizes the DMER methods as a baseline, which includes:
        (1) CRNN (Convolutional and Recurrent Neural Network) \cite{malik2017stacked}: This method stacks convolutional and recurrent neural networks to predict emotions using a compact architecture with fewer parameters.
        (2) DNN (Deep Neural Network) \cite{orjesek2019dnn}: This approach leverages convolutional and recurrent neural networks for feature extraction directly from raw audio without any preprocessing.
        (3) MCRNN (Multi-view Convolutional Recurrent Neural Network) \cite{he2020multi}: This technique employs multi-view CNNs and BiLSTM to automatically learn feature representations from raw audio, incorporating data augmentation methods.
        (4) DAMFF (Dual Attention-based Multi-scale Feature Fusion) \cite{zhang2023dual}: As the SOTA DMER method, it introduces a dual attention mechanism for temporal-frequency multi-scale feature fusion from spectrograms, employing BiLSTM for sequence learning.

        \subsubsection{Objective Metrics.} The performance of DSAML is evaluated using three objective metrics: Root Mean Square Error (RMSE), Pearson Correlation Coefficient (PCC), and Concordance Correlation Coefficient (CCC). Among these, the RMSE metric is used to measure the deviation between the predicted values and the actual values. A smaller RMSE value indicates higher prediction accuracy and lower error of the model. The PCC is used to assess the linear correlation between the predicted values and the actual values, with larger values indicating a stronger positive correlation. The CCC combines both precision and consistency aspects, serving as an improvement over the PCC. It not only considers the linear correlation but also takes into account the agreement between the means and variances of the predicted and actual values. A higher CCC value indicates better predictive performance of the model.

        \begin{table*}[!h]
        \centering
        \small
        \begin{tabular}{lcccccccccccc}
        \toprule
        \multirow{3}{*}{Model} & \multicolumn{6}{c}{DEAM} & \multicolumn{6}{c}{PMEmo} \\
        \cmidrule(lr){2-7} \cmidrule(lr){8-13}
            & \multicolumn{3}{c}{Arousal} & \multicolumn{3}{c}{Valence} & \multicolumn{3}{c}{Arousal} & \multicolumn{3}{c}{Valence} \\
        \cmidrule(lr){2-4} \cmidrule(lr){5-7} \cmidrule(lr){8-10} \cmidrule(lr){11-13}
        & CCC $\uparrow$ & PCC $\uparrow$ & RMSE $\downarrow$ & CCC $\uparrow$ & PCC $\uparrow$ & RMSE $\downarrow$ & CCC $\uparrow$ & PCC $\uparrow$ & RMSE $\downarrow$ & CCC $\uparrow$ & PCC $\uparrow$ & RMSE $\downarrow$ \\
        \midrule

        Ours & \textbf{0.39} & \textbf{0.675} & \textbf{0.202} & \textbf{0.104} & \textbf{0.221} & \textbf{0.26} & \textbf{0.186} & \textbf{0.386} & \textbf{0.112} & \textbf{0.108} & 0.26 & \textbf{0.122 }\\

        DAMFF & 0.354 & 0.621 & 0.217 & 0.055 & 0.043 & 0.299 & 0.175 & 0.378 & 0.118 & 0.091 & \textbf{0.335} & 0.134 \\

        MCRNN & 0.227 & 0.374 & 0.265 & 0.029 & -0.005 & 0.297 & 0.104 & 0.266 & 0.168 & 0.038 & 0.154 & 0.169 \\

        DNN & 0.188 & 0.362 & 0.261 & 0.033 & 0.003 & 0.281 & 0.087 & 0.217 & 0.151 & 0.03 & 0.134 & 0.159 \\

        CRNN & 0.244 & 0.517 & 0.218 & 0.017 & -0.039 & 0.27 & 0.076 & 0.231 & 0.183 & 0.034 & 0.224 & 0.169 \\

        \bottomrule
        \end{tabular}
        \caption{Performance of different models on two datasets of the traditional DMER task.}
        \label{tab:model_performance_non_personalized_task}
        \end{table*}

    \subsection{Objective Experiment}
    This paper evaluates the performance of DSAML on PDMER and traditional DMER separately. For PDMER, the test set uses personalized annotations from all annotators as labels, referred to as the personalized task. In contrast, for traditional DMER, we follow conventional validation methods where the test set uses the mean of annotations from all annotators as labels, referred to as the traditional DMER task.

    \subsubsection{Traditional DMER Task}
    In traditional DMER tasks, this paper follows the approach of other DMER studies by using the mean of all annotators as the label for validation. Table \ref{tab:model_performance_non_personalized_task} presents the performance of our method compared to other methods on the DEAM and PMEmo datasets.
    It is observed that our method outperforms other methods across all metrics on the DEAM dataset, and also performs well on the PMEmo dataset. This is mainly because the emotional state at any moment in DMER may be influenced by distant moments, and our model architecture design with a dual-scale attention transformer can better capture long-term dependencies, thereby improving prediction accuracy. This is also validated in the ablation study.
    Moreover, the PCC value in the valence dimension of PMEmo dataset is lower than the DAMFF method, which may be due to the shorter music segments in this dataset, with fewer long-term dependencies. The trend of the valence dimension can be more easily predicted by DAMFF based on short-term dependencies, whereas our model is better at capturing long-term dependencies.
    However, our method still outperforms other methods on other metrics of PMEmo dataset, especially the more comprehensive CCC metric, demonstrating that our method is more effective in traditional DMER tasks.

    \subsubsection{PDMER Task}
    \begin{table}[!h]
    \small
    \centering
    \begin{tabular}{lcccccc}
    \toprule
    \multirow{2}{*}{Model} & \multicolumn{3}{c}{Arousal} & \multicolumn{3}{c}{Valence} \\
    \cmidrule(lr){2-4} \cmidrule(lr){5-7}
     & CCC $\uparrow$ & PCC $\uparrow$ & RMSE $\downarrow$ & CCC $\uparrow$ & PCC $\uparrow$ & RMSE $\downarrow$ \\
    \midrule

    Ours & \textbf{ 0.377} & \textbf{0.541} & \textbf{0.236} & \textbf{0.092} & \textbf{0.154} & \textbf{0.245} \\
    DAMFF$^{\dagger}$ & 0.22 & 0.414 & 0.319 & 0.035 & 0.069 & 0.344 \\
    DAMFF & 0.232 & 0.478 & 0.342 & 0.035 & 0.008 & 0.39 \\
    MCRNN$^{\dagger}$ & 0.173 & 0.312 & 0.332 & 0.022 & 0.021 & 0.348 \\
    MCRNN & 0.176 & 0.301 & 0.364 & 0.015 & -0.002 & 0.391 \\
    DNN$^{\dagger}$ & 0.138 & 0.273 & 0.35 & 0.01 & -0.009 & 0.367 \\
    DNN & 0.153 & 0.286 & 0.361 & 0.016 & -0.004 & 0.385 \\
    CRNN$^{\dagger}$ & 0.113 & 0.29 & 0.31 & 0.008 & -0.009 & 0.337 \\
    CRNN & 0.131 & 0.365 & 0.35 & -0.001 & -0.021 & 0.366 \\

    \bottomrule
    \end{tabular}
    $^{\dagger}$ The models trained with MAML.
    \caption{Performance of different models in the PDMER task.}
    \label{tab:model_performance_personalized_task}
    \end{table}

    In personalized tasks, the test set requires annotation data from each annotator. Since PMEmo dataset does not include these data, so personalized tasks are validated only on the DEAM dataset. Table \ref{tab:model_performance_personalized_task} presents the performance of all methods in personalized tasks. Compared to other baseline methods, our method demonstrates superior performance across all metrics.
    This can be attributed to the effectiveness of our personalized strategy.
    Our personalized strategy retains personalized information in the dataset, and our task construction strategy enables the model to differentiate between different personalized perceptions.
    When facing new personalized tasks, our method can effectively utilize previous knowledge, thereby improving model performance in the PDMER task.
    Notably, incorporating MAML reduces the performance of other methods, as they learn the support set’s personality during training but predict the query set with different personalized perceptions during the training phase, which makes it difficult for the models to fit.

    \subsection{Subjective Experiment}
    To further demonstrate the effectiveness of DSAML for the PDMER task in real-world scenarios, we conducted a subjective user experiment to evaluate the accuracy of personalized emotion prediction.
    The experiment involved 22 participants, including 11 females and 11 males. Participants were first required to listen to a 45-second song and adjust the initial V-A curve generated by the DSAML to best match their perception. Based on this adjustment, participants then listened to 10 songs sequentially, all of which come from the test set of the DEAM dataset.
    After listening to each song, they were asked to rank four V-A curves by perceptual match.
    The four V-A curves were generated by the following methods: (1) Ground truth, (2) DSAML, (3) DSAML without personalized strategy, and (4) DAMFF.
    The ranking was based on the consistency between the V-A curve and the participant's perception, with 1 indicating the most consistency and 4 indicating the least consistency.

    \begin{figure}[t]
    \centering
    \includegraphics[width=\columnwidth]{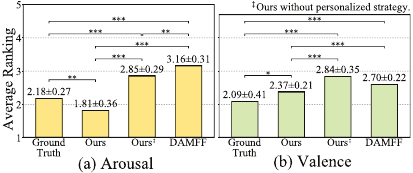}

    \caption{Average ranking of users using different methods in subjective experiment (*: $p < .05$, **: $p < .01$, ***: $p < .001$).}
    \label{fig:user-study}
    \end{figure}

    \begin{table*}[!h]
    \centering
    \small
    \begin{tabular}{lcccccccccccc}
    \toprule
    \multirow{3}{*}{Model} & \multicolumn{6}{c}{DEAM} & \multicolumn{6}{c}{PMEmo} \\
    \cmidrule(lr){2-7} \cmidrule(lr){8-13}
        & \multicolumn{3}{c}{Arousal} & \multicolumn{3}{c}{Valence} & \multicolumn{3}{c}{Arousal} & \multicolumn{3}{c}{Valence} \\
    \cmidrule(lr){2-4} \cmidrule(lr){5-7} \cmidrule(lr){8-10} \cmidrule(lr){11-13}
    & CCC $\uparrow$ & PCC $\uparrow$ & RMSE $\downarrow$ & CCC $\uparrow$ & PCC $\uparrow$ & RMSE $\downarrow$ & CCC $\uparrow$ & PCC $\uparrow$ & RMSE $\downarrow$ & CCC $\uparrow$ & PCC $\uparrow$ & RMSE $\downarrow$ \\
    \midrule
    Ours & \textbf{0.402} & \textbf{0.633} & \textbf{0.196} & \textbf{0.117} & 0.182 & \textbf{0.267} & \textbf{0.186} & \textbf{0.386} & 0.112 & \textbf{0.108} & \textbf{0.26} & \textbf{0.122} \\
    Ours w.o. Local-Attention & 0.385 & 0.624 & 0.215 & 0.077 & 0.134 & 0.268 & 0.167 & 0.341 & \textbf{0.112} & 0.087 & 0.213 & 0.123 \\
    Ours w.o. Global-Attention & 0.37 & 0.594 & 0.207 & 0.084 & 0.148 & 0.283 & 0.157 & 0.352 & 0.123 & 0.075 & 0.214 & 0.132 \\
    Ours w.o. Attention-Loss & 0.338 & 0.564 & 0.221 & 0.103 & \textbf{0.185} & 0.279 & 0.148 & 0.353 & 0.135 & 0.098 & 0.254 & 0.139 \\
    Ours w.o. Adapter & 0.329 & 0.529 & 0.221 & 0.083 & 0.12 & 0.272 & 0.138 & 0.318 & 0.121 & 0.077 & 0.247 & 0.125  \\
    \bottomrule
    \end{tabular}
    \caption{Performance of ours on two datasets of the traditional DMER task.}
    \label{tab:ablation_study_non_personalized_task}
    \end{table*}

    As shown in Figure \ref{fig:user-study}, we analyzed the consistency ranking of different methods in both arousal and valence values, and conducted a significance analysis using the paired t-test.
    In the arousal dimension, the average ranking of DSAML is at the forefront, and it has a high level of significance compared to other methods, indicating that our model can predict the most consistent personalized emotion with the participants' perceptions.
    More notably, our model even outperforms the ground truth of the dataset, which to some extent indicates the importance of personalization in the real world for the DMER task.
    In the valence dimension, the average ranking of our method is also significantly higher than that of the other methods except the ground truth.  %
    This may be due to valence's greater complexity, making it harder to predict than arousal. As Chua et al. pointed out, perceptions of arousal are primarily influenced by auditory information, while perceptions of valence can be influenced by both visual and auditory information \cite{chua2022predicting}. This indicates that the valence dimension is more challenging to predict using only audio information, which is consistent with our results in both objective and subjective experiments.

    \subsection{Ablation Study}
    In the ablation study, we evaluated the effectiveness of each component of our model. We conducted ablation studies on both traditional DMER and PDMER tasks.

    \subsubsection{Traditional DMER Task}

    Table \ref{tab:ablation_study_non_personalized_task} presents the performance of our model with different components removed on the DEAM and PMEmo datasets. It can be observed that the overall performance of our model drops when the local attention, global attention, attention loss and Imagebind adapter are removed, indicating that these components are essential for the model's performance. Moreover, the attention loss and Imagebind components have a more significant impact on the model's performance, which is consistent with our design concept.
    The attention loss component ensures that the local and global attention focuses on different scales, while the Imagebind adapter enables the model can extract local features which ignoring by Imagebind.

    \subsubsection{PDMER Task}
    Table \ref{tab:ablation_study_personalized_task} presents the performance of our model with different components removed in the PDMER task. We can observe that the overall performance drops when the MAML or personalized task construction strategy is removed, indicating that these components are essential for the model's performance. However, it is worth noting that the performance of the model without MAML but with the personalized task construction strategy is worse than the model without both components. This is because the personalized task construction strategy alone leads to multiple different labels for the same sample in the training set, making it difficult for the model to fit during training. This also explains why other traditional DMER methods often train the model using mean labels, as using the original labels directly will lead to a decrease in model performance.

    \begin{table}[!h]
    \small
    \centering
    \begin{tabular}{lcccccc}
    \toprule
    \multirow{2}{*}{Model} & \multicolumn{3}{c}{Arousal} & \multicolumn{3}{c}{Valence} \\
    \cmidrule(lr){2-4} \cmidrule(lr){5-7}
        & CCC $\uparrow$ & PCC $\uparrow$ & RMSE $\downarrow$ & CCC $\uparrow$ & PCC $\uparrow$ & RMSE $\downarrow$ \\
    \midrule

    Ours & \textbf{ 0.377} & 0.541 & \textbf{0.236} & \textbf{0.092} & \textbf{0.154} & \textbf{0.245} \\
    \makecell[l]{Ours $^\dagger$} & 0.305 & 0.518 & 0.332 & 0.062 & 0.095 & 0.383 \\
    \makecell[l]{Ours $^\ddagger$} & 0.357 & 0.533 & 0.236 & 0.08 & 0.129 & 0.256 \\
    \makecell[l]{Ours $^\dagger$$^\ddagger$} & 0.332 & \textbf{0.566} & 0.309 & 0.061 & 0.132 & 0.359 \\

    \bottomrule
    \end{tabular}
    \\{$^\dagger$ The models trained w.o. MAML. \\ $^\ddagger$ The models trained w.o. personalized task construction strategy.}
    \caption{Performance of ours in the PDMER task.}
    \label{tab:ablation_study_personalized_task}
    \end{table}

\section{Conclusion}
    This paper proposes the DSAML method for personalized dynamic music emotion recognition (PDMER).
    DSAML fuses features from a dual-scale feature extractor and captures both short and long-term dependencies using a dual-scale attention transformer, improving the performance in traditional DMER.
    Moreover, a personalized strategy is proposed, which apply a novel task construction strategy into the MAML training process. The proposed task construction strategy divides tasks by annotators, ensuring consistent perception. Leveraging this strategy alongside meta-learning, DSAML can predict personalized perception of emotions with just one personalized annotation sample.
    Objective experimental results demonstrate that DSAML outperforms previous music emotion recognition methods in both traditional DMER and PDMER tasks. Furthermore, subjective experiments validate the effectiveness of DSAML in real-world scenarios.

\section{Acknowledgments}

This work is supported by the National Key Research and Development Program of China (2023YFF0904900), and the National Natural Science Foundation of China (No. 62272409).

\bibliography{aaai25}

\end{document}